\documentclass[10pt,twocolumn]{article} 
\usepackage{simpleConference}
\usepackage{times}

\pdfoutput=1

\usepackage{graphicx}
\usepackage{amssymb}
\usepackage{url}

\usepackage{float} 

\usepackage{multirow}
\usepackage{siunitx}

\usepackage{booktabs,caption}
\usepackage[flushleft]{threeparttable}

\usepackage{dcolumn}   
\usepackage{bm}        
\usepackage{amssymb}   
\usepackage{float} 
\usepackage{amsmath}   

\usepackage{rotating}
\usepackage{booktabs}

\usepackage{pdflscape}
\usepackage{afterpage}
\usepackage{capt-of}

\usepackage{esvect}
\usepackage{epstopdf}
\usepackage{bm}        
\usepackage{amssymb}   
\usepackage{amsmath}   
\makeatletter
\newcommand{\mathleft}{\@fleqntrue\@mathmargin20pt}
\newcommand{\mathcenter}{\@fleqnfalse}
\makeatother

\makeatletter
\def\blfootnote{\xdef\@thefnmark{}\@footnotetext}
\makeatother
\begin{document}

\vspace{2cm}

\title{Multipole sensitivity to phase variation in pion photo-and electroproduction analyses}

\author{L. Markou$^{1}$, E. Stiliaris$^{2}$ and C. N.Papanicolas$^{1,*}$  \\
\\
$^{1}$ The Cyprus Institute, K. Kavafi 20, 2121 Nicosia, Cyprus \\ 
$^{2}$ National and Kapodistrian University of Athens, Physics Department, 15771 Athens, Greece \\
\today
\\
\\
}


\twocolumn[
  \begin{@twocolumnfalse}
    \maketitle
    \begin{abstract}
   We use the Athens Model Independent Analysis Scheme (AMIAS) 
to examine the validity of using the Fermi-Watson theorem in  the multipole analyses of pion photoproduction 
and electroproduction data.  
A standard practice in this field is to fix the multipoles' phases from $\pi N$ scattering data, 
making use of the Fermi - Watson theorem. However, 
these phases are known with limited accuracy and the effect of this
uncertainty on the obtained multipole extraction has not been fully explored yet. Using AMIAS we 
constrain the phases  within their experimentally determined uncertainty. 
We first analyze  sets of pseudodata of increasing statistical precision and subsequently  we 
apply the methodology for a re-analysis of the Bates/Mainz electroproduction data.
It is found that the uncertainty induced by the $\pi N$ phases uncertainty to the extracted solutions 
 would be significant  only in the analysis of data with much higher 
precision than the current available experimental data.  
\end{abstract}

\\
\\

  \end{@twocolumnfalse}
]

\thispagestyle{empty}

\section{Introduction}
\label{intro}
\blfootnote{$^{*}$ Corresponding Author: cnp@cyi.ac.cy} Compton scattering, pion photoproduction,   
and pion - nucleon scattering are related by unitarity through a common
S matrix \cite{blanpied2001n} and the Fermi-Watson (FW) \cite{FW} theorem requires the $( \gamma , \pi )$ and $( \pi , \pi )$ 
channels to have the same phase
 below the two-pion threshold. Multipole analyses below this  threshold  are subject to this theoretical constraint 
 which requires all multipoles 
with different character  but the same quantum numbers $I, l , J$ to have the same phase $\pm n \pi$ 
which is the same as the corresponding $\pi N$ scattering phase shift.
The  pion photoproduction multipole phases and  the scattering phase shifts are related through \cite{FW}:
\begin{equation}
\label{eq:fw}
A^{I}_{l\pm} = |A^{I}_{l\pm}|e^{i \left ( \delta_{IlJ} + n \pi \right )} 
\end{equation}
where $\delta_{IlJ}$ is the pion - nucleon scattering phase shift, $I$ is the isospin quantum number, $l$ the angular momentum, 
$J$ the total angular momentum and 
"$\pm$" is used to distinguish whether  $J$ and the spin are parallel or anti-parallel. 
 $A^{I}_{l}=\{E^{I}_{l},M^{I}_{l},L^{I}_{l}\}$ denotes the electric, magnetic or longitudinal nature of the multipole.
 As $\pi N$ scattering phase shifts are easier to measure and therefore are known with higher precision, 
 this theoretical constraint provides a very powerful tool in photoproduction (and electroproduction) multipole 
 analyses. Multipoles are complex functions of the center mass energy $W$ 
and by applying the FW theorem  the number of  unknown parameters is halved since only the moduli of the multipoles 
$|A_{l\pm}^{I}|$ needs to be determined. It has been widely used in multipole analyses of both 
pion photoproduction data \cite{blanpied1997n,PhysRevLett.78.606,beck2000determination,kotulla2007real,markou1} and pion electroproduction data 
\cite{PhysRevLett.82.45,PhysRevLett.86.2963,sparveris2005investigation,stave2006lowest}. 

The values  of the $\pi N$ scattering phase shifts are  known from the analyses of $\pi N$ 
scattering data, \textit{e.g.} ref. \cite{workman2012parameterization}. 
 The FW applies 
 well beyond the two pion threshold as  the $\pi N$ inelasticities  are very small \cite{beck2000determination,pseudo}. 
For example, in the  pion photoproduction data analysis by Grushin \cite{grushin} where both the real and imaginary parts of 
the $\ell \leq 1$ multipoles 
 were determined without using the FW theorem it was found that the mean difference  between the $\delta_{33}^{\gamma,\pi N}$ and the 
  pion - nucleon scattering phase shift was only $-(2.3 \pm 0.5)^{\circ}$ over the energy range $E_{\gamma}^{lab} = 250 - 500$ $MeV$.
%
%

\begin{figure}[htp]
 \centering
 {\includegraphics[width = 3.4in]{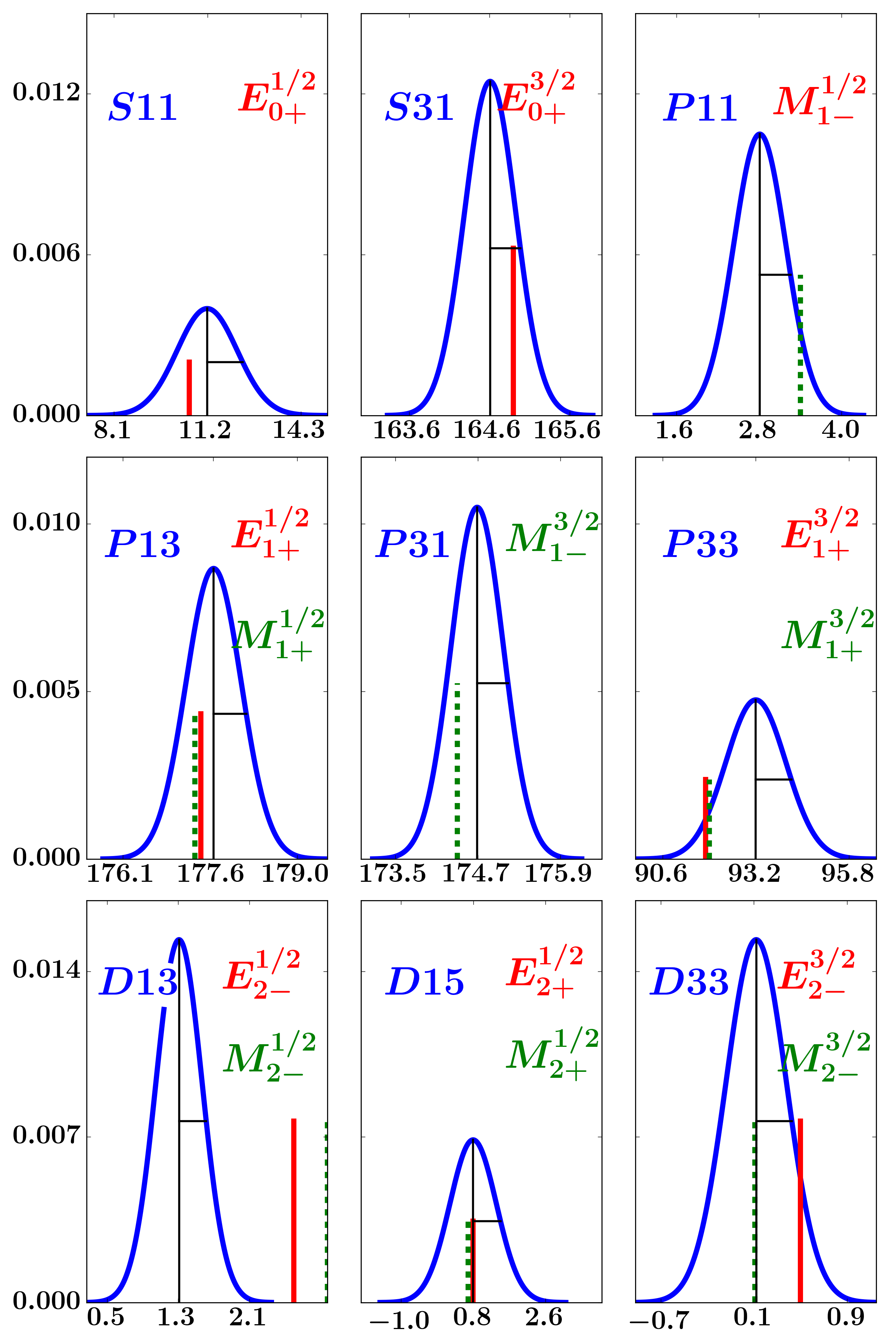}}\hfil
\caption{Plotted as normal distributions with $N[\mu,2\sigma]$ are the nine reported pion-nucleon scattering phase shifts  of the 
single energy analysis WI08 \cite{gwuweb2} 
at $W_{cm}=1234.5$ $MeV$. Vertical lines show the  MAID07 model values \cite{Drechsel:2007if} for the corresponding  
pion  photoproduction multipole phases ($\pm n\pi$), coded as red-continuous line for the Electric multipoles and green-dotted line for the Magnetic. 
 The distributions  are normalized to unity and are given in degrees.}
\label{fig:piN_phases_2sigma}
\end{figure}

Fig. \ref{fig:piN_phases_2sigma} shows the nine $\pi N$ phases reported in ref. \cite{workman2012parameterization} 
as part of the WI08  
 partial wave analysis  at $W=1235$ $MeV$ at the photon point.  The exact numerical values are available online \cite{gwuweb2}. 
 Each phase is plotted as a Gaussian, $N[\mu,2\sigma]$, 
with the same mean value ($\mu$)  and double the statistical uncertainty ($\sigma$) derived from the experimental $\pi N$ data. 
The MAID07 model prediction \cite{Drechsel:2007if} 
for the corresponding  pion  photoproduction multipole phases ($\pm n\pi$), at the same kinematics, is also shown.
The $\pi N$ phases are known with limited accuracy, and although multipole analyses use them as if they are known with 
 infinite precision,  the effect of this
uncertainty on the extracted multipoles has not been  explored yet. Using the Athens Model Independent Analysis Scheme (AMIAS) 
we 
achieve this by constraining the phases  within their experimentally determined uncertainty,  making the analysis 
Bayessian.

The following  sections are organized as follows: 
In Sec. \ref{sec:fw_methodology} we discuss the methodology for multipole extraction with the AMIAS and 
the inclusion  of parameters with known uncertainties. In Sec. \ref{sec:fw_pseudo} we detail the 
creation of pion photoproduction pseudodata of predetermined statistical precision. The multipole content of those 
pseudodata is derived in Sec. \ref{sec:fw_results} validating the methodology described in Sec. \ref{sec:fw_methodology}. 
 In Sec. \ref{sec:fw_bates} we apply the same methodology for a re-analysis of the Bates/Mainz electroproduction data 
 \cite{sparveris2005investigation} 
  measured at  $Q^2=0.127$ $GeV^2/c^2$ and $W=1232$ $MeV$.  Concluding remarks are given in Sec. \ref{sec:fw_conclusions}.  

%

\section{Methodology}
\label{sec:fw_methodology}
The methodology employed is the implementation of the  Chew, Goldenberg, Low and Nambu (CGLN) theoretical framework  
  \cite{chew1957relativistic} for single energy multipole analyses in the Athens Model Independent Analysis Scheme (AMIAS) 
\cite{stiliaris2007multipole,papanicolas2012novel}. 
The AMIAS method is based on statistical concepts and relies heavily on Monte Carlo and simulation techniques, 
and it thus requires High Performance Computing as it is
computationally intensive. The method identifies and determines with maximal 
precision parameters that are sensitive to the data by yielding their Probability Distribution Functions (PDF). 
The AMIAS is computationally robust and numerically stable. It has been successfully applied in the analysis of 
data from nucleon photo-and electroproduction resonance \cite{markou1,stiliaris2007multipole,RevModPhys.84.1231}, 
lattice QCD simulations \cite{alexandrou2015novel} and medical imaging \cite{loizospseudo}.

AMIAS requires that the parameters 
to be extracted from the experimental data are explicitly linked via a theory or a model \cite{stiliaris2007multipole}. 
In the case of pion photoproduction this requirement is  provided by the CGLN theory  as in ref.  \cite{markou1} and in the case of  
electroproduction as in ref. \cite{stiliaris2007multipole}.  
 The  multipoles are connected to the pion photoproduction observables via the CGLN \cite{chew1957relativistic} 
 amplitudes $\left ( F_i, i=1,6 \right ) $:
\mathleft
\begin{equation}  \label{eq:cgln1}
 \begin{split}
 F_{1}  = {} &  \sum_{l=0}^{\infty}   [ \left ( lM_{l+}  + E_{l+} \right ) P'_{l+1}(x)   \\
  & + \left ( \left (l+1 \right ) M_{l-} + E_{l-} \right ) P'_{l-1}(x) ]
\end{split}
\end{equation}
\begin{equation}
  \label{eq:cgln2}
 F_{2}  = {} \sum_{l=1}^{\infty} \left [\left (l+1 \right)  M_{l+}  + lM_{l-}  \right ]  P'_{l}(x)
\end{equation}
\begin{equation}   \label{eq:cgln3}
 \begin{split}
  F_{3}  = {} & \sum_{l=1}^{\infty}  [ \left ( E_{l+} - M_{l+}  \right ) P''_{l+1}(x) \\ 
  & + \left ( E_{l-}  + M_{l-}  \right ) P''_{l-1}(x)  ]   
\end{split}
\end{equation}
\begin{equation}
 F_{4}  = {} \sum_{l=2}^{\infty} \left [ M_{l+}  - E_{l+}  - M_{l-}  - E_{l-} \right ] P''_{l}(x)  
 \label{eq:cgln4}
\end{equation}
\begin{equation}
 F_{5}  = {} \sum_{l=0}^{\infty} \left [ (l+1)L_{l+} P'_{l+1}(x)-lL_{l-}P'_{l-1}(x) \right ]   
 \label{eq:cgln5}
\end{equation}
\begin{equation}
 F_{6}  = {} \sum_{l=1}^{\infty} \left [ l L_{l-} - (l+1) L_{l+}  \right ] P'_{l}(x)  
 \label{eq:cgln6}
\end{equation}
where $x = \cos(\theta)$ is the cosine of the scattering angle and $P'_{l}$ are the derivatives of the Legendre polynomials. 
Multipoles $A_{l\pm}=\{E_{l\pm}, M_{l\pm}, L_{l\pm} \}$ refer to the electric, magnetic or longitudinal nature of the photon respectively.
At the real photon point, longitudinal degrees of freedom in the photon's polarization  vanish identically 
and the $\gamma N \to \pi N$ reaction  is described solely by  the CGLN amplitudes $F_{1}$ to $F_{4}$.

From isospin conservation in the pion-nucleon system it follows that the multipoles can
be expressed in terms of definite isospin  \cite{BERENDS19671,dreschsel1992threshold}, namely, the $A^{1/2}$ and $A^{3/2}$ 
multipoles. These  are obtained from the reaction channel multipoles and the relations \cite{dreschsel1992threshold}:
\mathcenter
\begin{align}
A^{1/2} &={} \frac{A_{p \pi^{0}}}{3} +  \frac{\sqrt{2} A_{n \pi^{+}}}{3}, & A^{3/2}  &={}A_{p \pi^{0}} - \frac{A_{n \pi^{+}}} {\sqrt{2}}
\label{eq:epjaiso}
\end{align}
In contrast to the standard practice adhered up to now  
where the multipole  phases are considered as if known with infinite 
precision \cite{beck2000determination,sparveris2005investigation} 
and therefore treated as fixed parameters of the problem we 
 allow those phases to vary within their experimentally determined 
uncertainty obtained from $\pi N$ experiments. This  allows the  prior knowledge 
on  the multipole phases to be incorporated in the analysis. 

To ascertain the magnitude of the effect this phase variation induces on the  derived multipoles we examine   
 three sets of pseudodata where each set was created with  predetermined and increasing  statistical precision. 
For each pseudodata set three multipole analyses were performed differentiated by the manner in which the multipole phases were treated; 
 during  the first analysis  phases were fixed to the  values of the generating model,   
 during the second analysis  phases were fixed to the SAID-WI08 \cite{workman2012parameterization,gwuweb2}  
 model dependent analysis values and during the third 
 multipole phases were allowed to vary with Gaussian weight, with mean value and twice the standard deviation of that reported 
 by the SAID-WI08 single energy solution \cite{workman2012parameterization,gwuweb2}. In  implementing the phase variation,  
 and according to eq. \ref{eq:fw}, we  imposed  that during the variation procedure all multipoles with the same 
quantum numbers $I,l,J$ had the exact same phase $\pm \pi$. 
In contrast, multipole phases with different quantum numbers were varied independently.  

%

\section{Creation  of pseudodata}
\label{sec:fw_pseudo}
\mathleft
\begin{table*}
\large{
\begin{equation*}
\begin{aligned}
d\sigma_{0} & =   \operatorname{Re} \left [ F_{1}^{*}F_{1} + F_{2}^{*}F_{2} + \sin^{2} \theta \left ( F_{3}^{*}F_{3}/2 + F_{4}^{*}F_{4}/2 + F_{2}^{*}F_{3} + F_{1}^{*}F_{4} + \cos \theta F_{3}^{*}F_{4} \right ) -2\cos \theta F_{1}^{*}F_{2} \right ] \rho \\
\hat{\Sigma}& =  -\sin^{2}\theta \operatorname{Re} \left [ \left ( F_{3}^{*}F_{3} + F_{4}^{*}F_{4} \right )/2 + F_{2}^{*}F_{3} + F_{1}^{*}F_{4} + \cos\theta F_{3}^{*}F_{4} \right ] \rho \\
\hat{T}     &=    \sin\theta \operatorname{Im} \left [  F_{1}^{*}F_{3} - F_{2}^{*}F_{4}  + \cos \theta \left (F_{1}^{*}F_{4} - F_{2}^{*}F_{3} \right ) - \sin^{2} \theta F_{3}^{*}F_{4} \right ] \rho \\
\hat{P}     &=   -\sin\theta \operatorname{Im} \left [ 2F_{1}^{*}F_{2} + F_{1}^{*}F_{3}  - F_{2}^{*}F_{4} - \cos \theta \left (F_{2}^{*}F_{3} - F_{1}^{*}F_{4} \right ) - sin^{2} \theta F_{3}^{*}F_{4} \right ] \rho \\
\hat{E}     &=   \operatorname{Re} \left [ F_{1}^{*}F_{1} + F_{2}^{*}F_{2}  - 2 \cos \theta F_{1}^{*}F_{2} + \sin^{2} \theta \left (F_{2}^{*}F_{3} + F_{1}^{*}F_{4} \right )  \right ] \rho \\
\hat{F}     &=   \sin \theta \operatorname{Re} \left [ F_{1}^{*}F_{3} - F_{2}^{*}F_{4}  -  \cos \theta \left ( F_{2}^{*}F_{3} - F_{1}^{*}F_{4}  \right )  \right ] \rho \\
\hat{G}     &=   \sin^{2} \theta \operatorname{Im} \left [ F_{2}^{*}F_{3} + F_{1}^{*}F_{4} \right ] \rho \\
\hat{H}     &=   \sin \theta \operatorname{Im} \left [ 2F_{1}^{*}F_{2} + F_{1}^{*}F_{3} - F_{2}^{*}F_{4} + \cos \theta \left ( F_{1}^{*}F_{4} - F_{2}^{*}F_{3} \right ) \right ] \rho \\
\end{aligned}
\end{equation*}
}
\caption{The CGLN content of the four single ($d\sigma_{0}$, $\hat{\Sigma}$, $\hat{T}$, $\hat{P}$) and four beam-target ($\hat{E}$, $\hat{F}$,$\hat{G}$, $\hat{H}$) polarization observables. 
The definitions $\rho = q/k$  and  $\hat{O} = O / d\sigma_0 $ are used. Angle $\theta$ is the center of mass scattering angle. }
\label{eq:Photo_kin_obs}
\end{table*}

\begin{figure}[htp]
\centering
{\includegraphics[width = 3.5in]{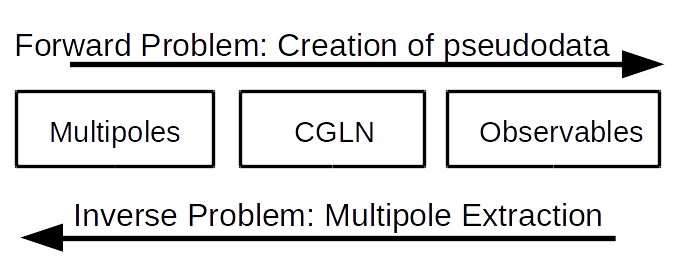}}\hfil
\vspace*{-5mm}
\caption{Multipole extraction in nucleon resonance photoproduction is an inverse problem in which the parameters to be extracted (multipoles) are connected to the experimental 
quantities via the CGLN formalism. To create  pseudodata a forward procedure is followed in which known multipole input is used to form the 
CGLN amplitudes and subsequently the photoproduction  observables.}
\label{fig:fw_inverse}
\end{figure}

We have created pseudodata for the four single $(d\sigma_0$, $\hat{\Sigma}$, $\hat{T}$, $\hat{P})$ and four double beam-target 
$(\hat{E}$, $\hat{F}$, $\hat{G}$, $\hat{H})$ polarization observables 
for the $\gamma p \to p \pi^0$ and $\gamma p \to n \pi^+$ reactions. The definitions used for the observables are the same as in ref. \cite{pseudo}. 
To create the pseudodata the MAID07 multipole solution at the photon point and at center mass energy $W=1234.5$ $MeV$  
 was inserted  in the CGLN multipole series, Eqs. 2-5, which were then used to construct 
 the photoproduction observables defined in Table \ref{eq:Photo_kin_obs}.  A schematic of this ``forward procedure'' is given 
 in Fig. \ref{fig:fw_inverse}. The observables  were subsequently  randomized according to the process:
\mathleft
\begin{equation}
 \label{eq:ran1}
 \begin{split}
 O_{i}^{k} =&{}  O_{i}^{k}  +  N[\mu,\sigma] \cdot  O_{i}^{k} \\ 
 \sigma_{O_{i}^{k}} =&{}   N[\mu,\sigma] \cdot O_{i}^{k}  
\end{split}
\end{equation}
where $O$ is the MAID07 model prediction,  $k$ distinguishes between each of the spin observables, $i$ labels  the angle, 
 $N[\mu,\sigma]$ is a normal distribution with known mean $(\mu)$ and standard deviation $\sigma$ and 
 $\sigma_{O_{i}^{k}}$ is the uncertainty attributed to the $i^{th}$ angular measurement of the $k^{th}$ observable. 
  
Using Eq. \ref{eq:ran1} we created 9000 sets of pseudodata. Each pseudodata set consisted
of $288$ datapoints; $18$ evenly spaced angular measurements in the dynamical region $\theta_{cm} \in[5^{\circ} : 175^{\circ} ]$ for each 
of the eight  polarization observables listed in Table  \ref{eq:Photo_kin_obs} for each proton target reaction. The angle $\theta_{cm}$ is defined as the angle between the incoming 
photon and the produced pion in the center of mass frame. 

\begin{figure}[htp]
\centering
{\includegraphics[width = 3.5in]{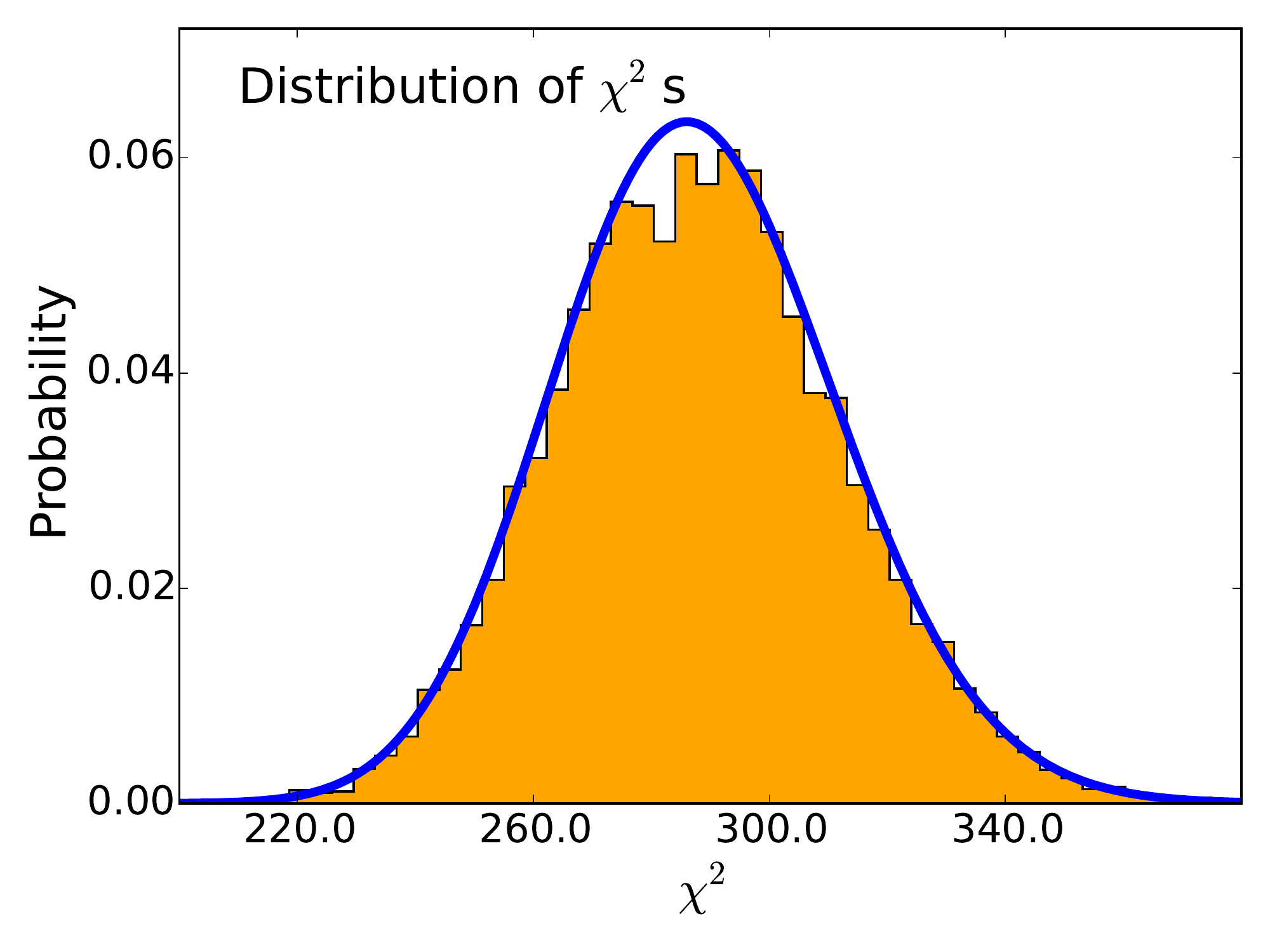}}\hfil
\vspace*{-5mm}
\caption{Histogram of the resulting $\chi^2$s as the generating model is passed without fitting through each dataset. 
Each dataset contains 288 datapoints. The histogram is described accurately by a  $\chi^2$ distribution of degrees of freedom equal to 
the number of datapoints of each set.}
\label{fig:fw_paper_distofchi2}
\end{figure}

The generated values and uncertainties of the pseudodata sets are shown to have the required behavior  \cite{friar1973determination,friar1975determination}. 
 by examining the resulting $\chi^2$ 
distribution. The $\chi^2$ distribution resulting by comparing each dataset to the generator, shown in Fig. \ref{fig:fw_paper_distofchi2}, 
is correctly described by the $\chi^2$ distribution with degrees of freedom equal to the number of datapoints $(=288)$ of the datasets. 

 

\begin{figure*}[htp]
\centering
{\includegraphics[width = \textwidth]{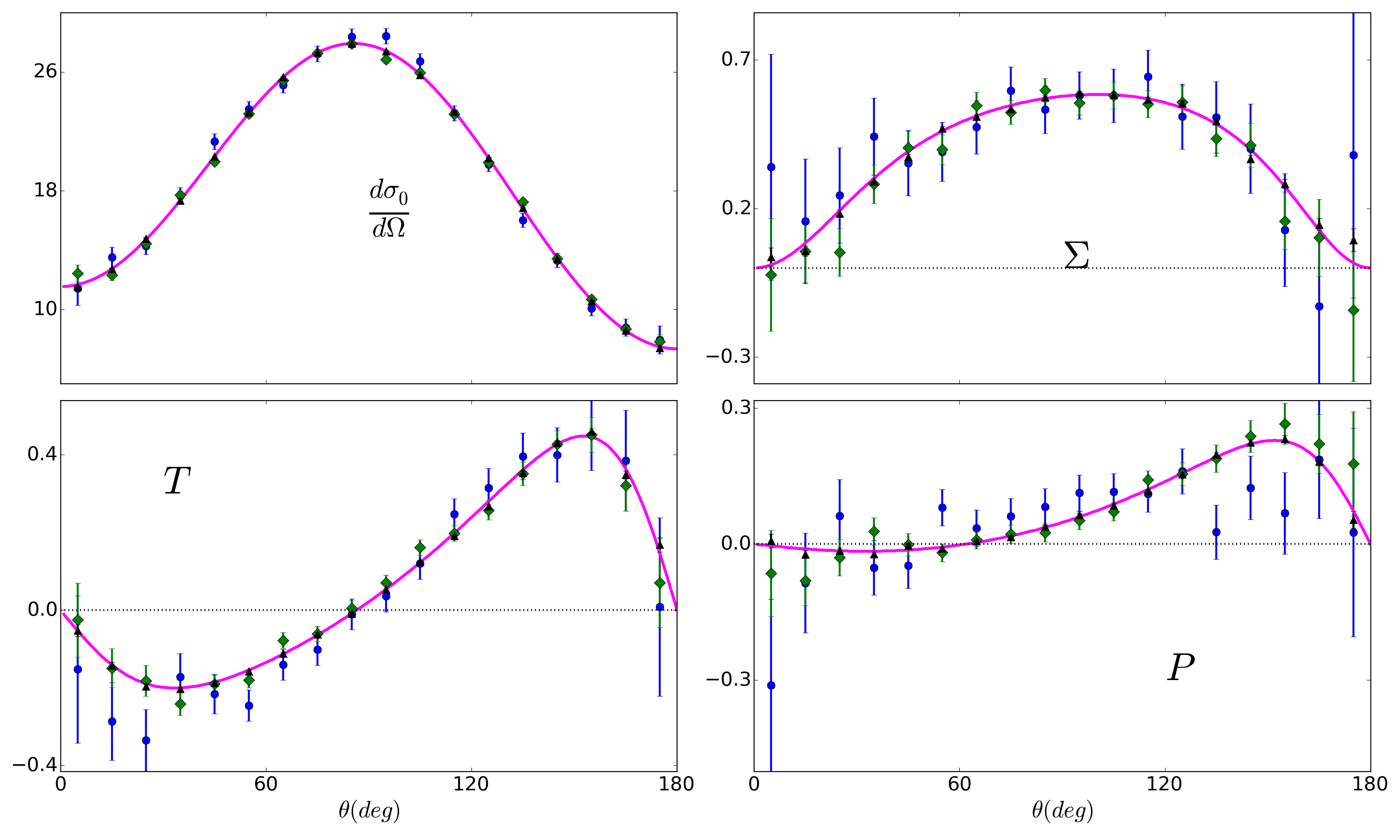}} \\
{\includegraphics[width = \textwidth]{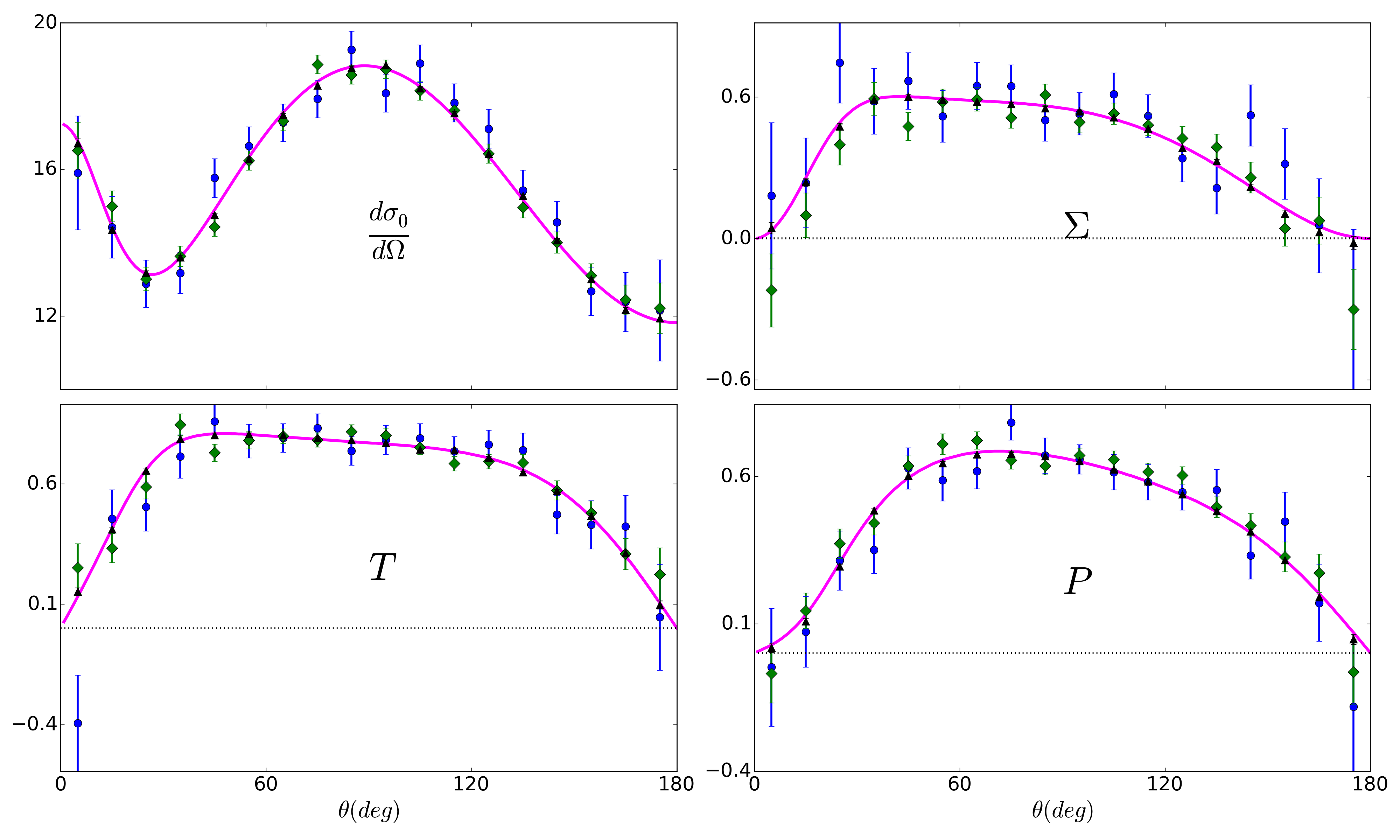}} \hfil
\vspace{-20pt}
\caption{Illustration of  the three sets of pseudodata with increasing precision for the differential cross section and the three single spin 
observables of the 
  $\gamma p \to p \pi^{0}$ reaction (top four panels) and the $\gamma p \to n \pi^{+}$ reaction (bottom four panels). 
  With blue circles are pseudodata of relative precision $2.0$, with green diamonds of relative precision $1.0$ 
   and with black triangles of relative precision $0.16$. The  precision of each dataset is given in relation to the precision of 
   current experimental measurements \cite{markou1,adlarson2015measurement}. 
   The continuous  magenta  curve is the MAID07 prediction (generator).  
   The differential cross sections are given in units of $10^{-3}/m_{\pi}$.}
\label{fig:fw_paper_ppiz_pseudo}
\end{figure*}

The methodology followed in creating the pseudodata through eq. \ref{eq:ran1} allows to control the precision of the generated pseudodata. 
Three distinct classes of pseudodata were created; each class featuring pseudodata of 
 statistical uncertainty  a) $2.0$, b) $1.0$ and c) $0.16$ times  
the statistical uncertainty  of the most precise pion photoproduction data  \cite{adlarson2015measurement} available to date. 
Fig. \ref{fig:fw_paper_ppiz_pseudo}  show pseudodata for the differential cross section ($d\sigma_0/d\Omega$), 
the beam asymmetry ($\hat{\Sigma}$), 
the target asymmetry ($\hat{T}$) and the recoil target asymmetry ($\hat{P}$) for the two proton target reactions.  
 The blue circles are used for  pseudodata of relative uncertainty $2.0$, the green diamonds for pseudodata of relative uncertainty $1.0$ 
   and the  black triangles for relative uncertainty $0.16$. The continuous magenta curve is the generator. The pseudodata 
   present some qualitative similarities  to   experimental data; the forward peak in the  $\gamma \pi \to n\pi^{+}$ differential cross section, 
   the absence of  such  peak in the $\gamma \pi \to p\pi^{0}$ differential cross section and  larger uncertainties 
   in the very forward and backward angles.  The generated pseudodata also provide spin observables which have never been measured before, e.g. 
   the beam-target $\hat{E}$, and full angular coverage.

%
%

\section{Results}
\label{sec:fw_results}
We applied the methodology presented in Sec. \ref{sec:fw_methodology} 
to the pseudodata and   we extracted values for all multipole amplitudes with relative angular momentum $\ell \leq 2$. 
Higher multipoles and up to all orders were frozen to the generating model values. 
The multipole amplitude PDFs derived from the AMIAS analyses were fitted with Gaussians and numerical results were extracted. 
Table \ref{tab:fw_paper_tab_moduli} lists the mean value  $\pm 34\%$ uncertainty ($1\sigma$) for the derived multipoles 
for two distinct analyses: 
Column ``MD07''  and ``W108''  refer to  analyses where the multipole phases were fixed to the MAID07 values (which is the generating model) or 
the WI08 solution respectively. 
Column ``Varied''  denotes  analyses where the  multipole phases were varied in a Gaussian manner with $N[\mu,2\sigma]$ where the mean value 
is taken from the WI08 $W$-dependent solution and the standard deviation, $\sigma$, the derived uncertainty of the WI08 single energy fit.

\begin{figure}[htp]
\centering
{\includegraphics[width = 3.6in]{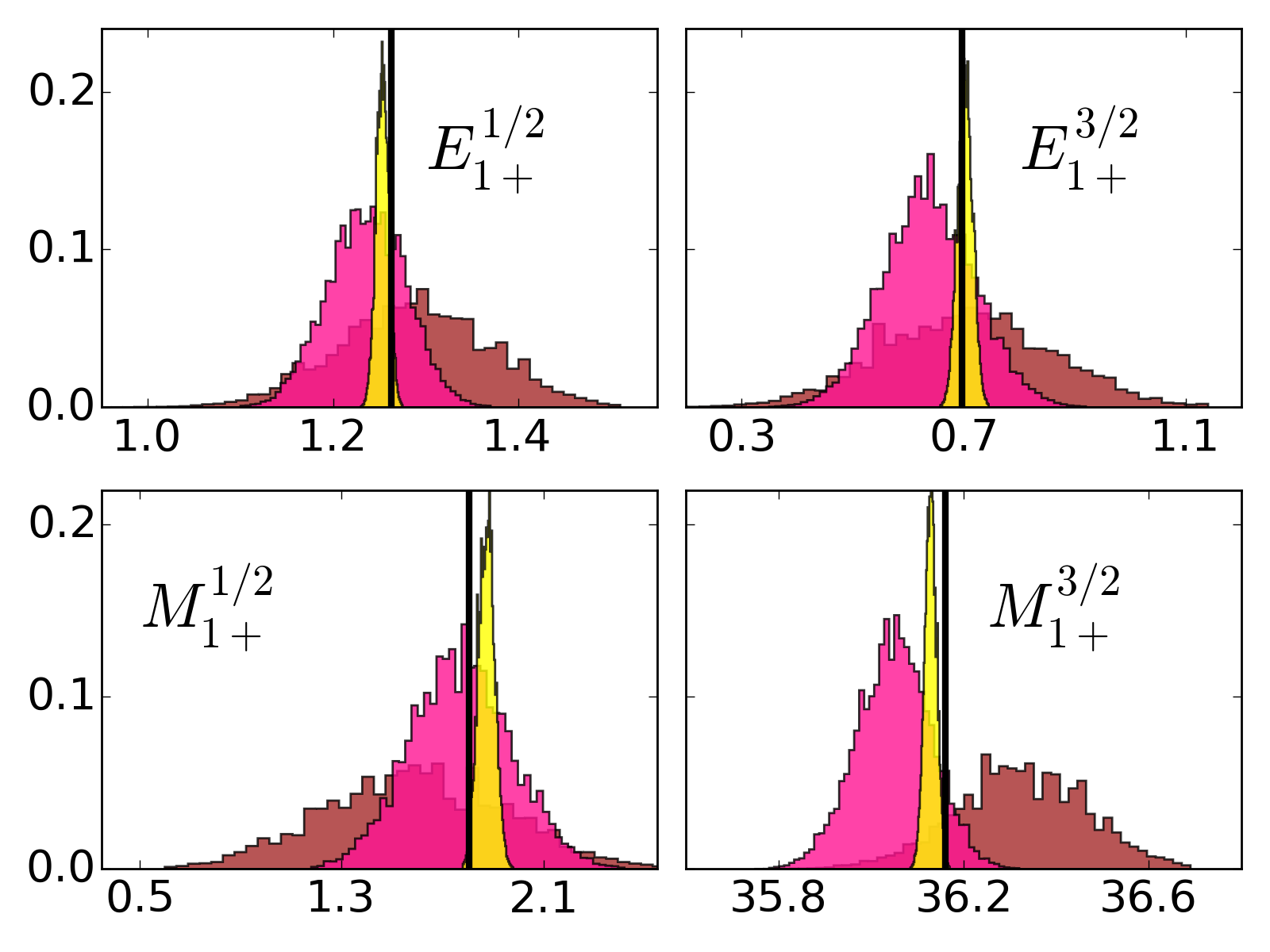}} \\[-1ex]
\caption{Probability Distribution Functions for the moduli of selected amplitudes derived from set A (brown), set B (magenta) and set C (yellow). 
The black vertical lines are the generator values (MAID07). Amplitudes are given in units of $10^{-3}/m_{\pi}$.}
\label{fig:fw_paper_amps_incr_prec}
\end{figure}

 The derived multipole values and uncertainties are  in good statistical agreement  
 with the generator input. 
 The  derived multipole uncertainty from each pseudodata set for the analyses with the phases fixed, listed as   ``MD07''  and ``W108''  
 in Table  \ref{tab:fw_paper_tab_moduli}, is reduced according to the statistical precision of each set. The pseudodata sets, Set A, Set B and 
 Set C were created with relative uncertainties $2.0$, $1.0$, and $0.16$ respectively. This is reflected in our results  as the 
 uncertainty associated with a specific multipole amplitude derived from Set A is reduced by a factor of $2.0$ and $12.5$ 
 when derived from Sets B and C  respectively. 
 This behavior indicates that the AMIAS method yields exact uncertainties 
 with a  precise statistical meaning \cite{stiliaris2007multipole,papanicolas2012novel}. 
 Fig. \ref{fig:fw_paper_amps_incr_prec} shows the PDFs of some selected amplitudes derived from the analysis of each  pseudodata set with the 
 multipole phases fixed to the MAID07 (generator) values. 
 As expected the derived uncertainty of each multipole amplitude is seen to decrease according 
 to the statistical precision of the analyzed pseudodata.

\begin{figure}[htp]
\centering
{\includegraphics[width = 3.6in]{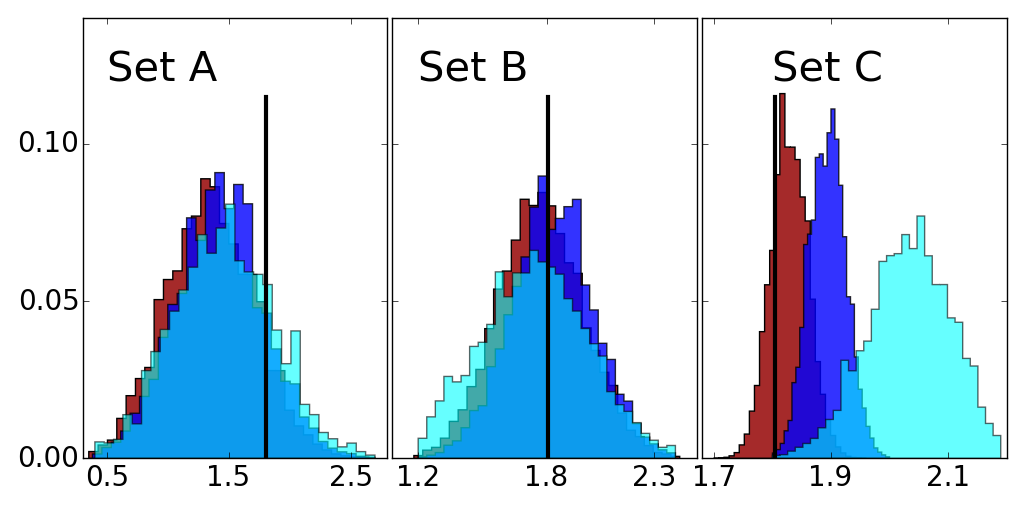}} \\[-1ex]
\caption{Probability Distribution Functions (PDFs) for the modulus of $M_{1+}^{1/2}$  derived from different pseudodata sets, each with 
increasing statistical precision. The PDFs are color coded according to the treatment of multipole phases during each analysis; 
brown (phases fixed to MAID07), blue (phases fixed to SAID-WI08) and cyan (phases varied with Gaussian weight. 
The black vertical lines is the generator value (MAID07). Amplitudes are given in units of $10^{-3}/m_{\pi}$.}
\label{fig:fw_paper_mp1rr1_allsets}
\end{figure}

Regarding the analyses of the pseudodata sets A and B, which are characterized by uncertainties greater or equal to current experimental data, 
the derived results are statistically equivalent whether the analysis was carried with the multipole  phases   
fixed to the generator values (MAID07), to the WI08 solution, or they were allowed to vary within the allowed experimental uncertainty. 
This is exhibited in Fig. \ref{fig:fw_paper_mp1rr1_allsets} for the case of the $M_{1+}^{1/2}$ amplitude. It demonstrates 
that  data of the currently available precision are not sensitive to such small changes or variations in the multipole phase.  
 The standard practice in multipole analyses, to treat these phases as if known with infinite precision, does not induce 
 additional  model bias to the derived multipoles. Regarding the analyses of set C we note  
 significant differences in the derived multipole mean values when phases change from the  MAID07 values to the  WI08 solution while the derived 
 uncertainty remains unchanged. When the  multipole phases are allowed to vary  the derived multipole mean values are shifted while their 
 associated uncertainty   is increased.  This increase is more prominent in 
 the background multipole  amplitudes $M_{1+}^{1/2}$ and $E_{0+}^{3/2}$.  Fig. \ref{fig:fw_paper_mp1rr1_allsets} shows this behavior 
 for the case of  the $M_{1+}^{1/2}$ amplitude.

\begin{figure}[htp]
\centering
{\includegraphics[width = 3.4in]{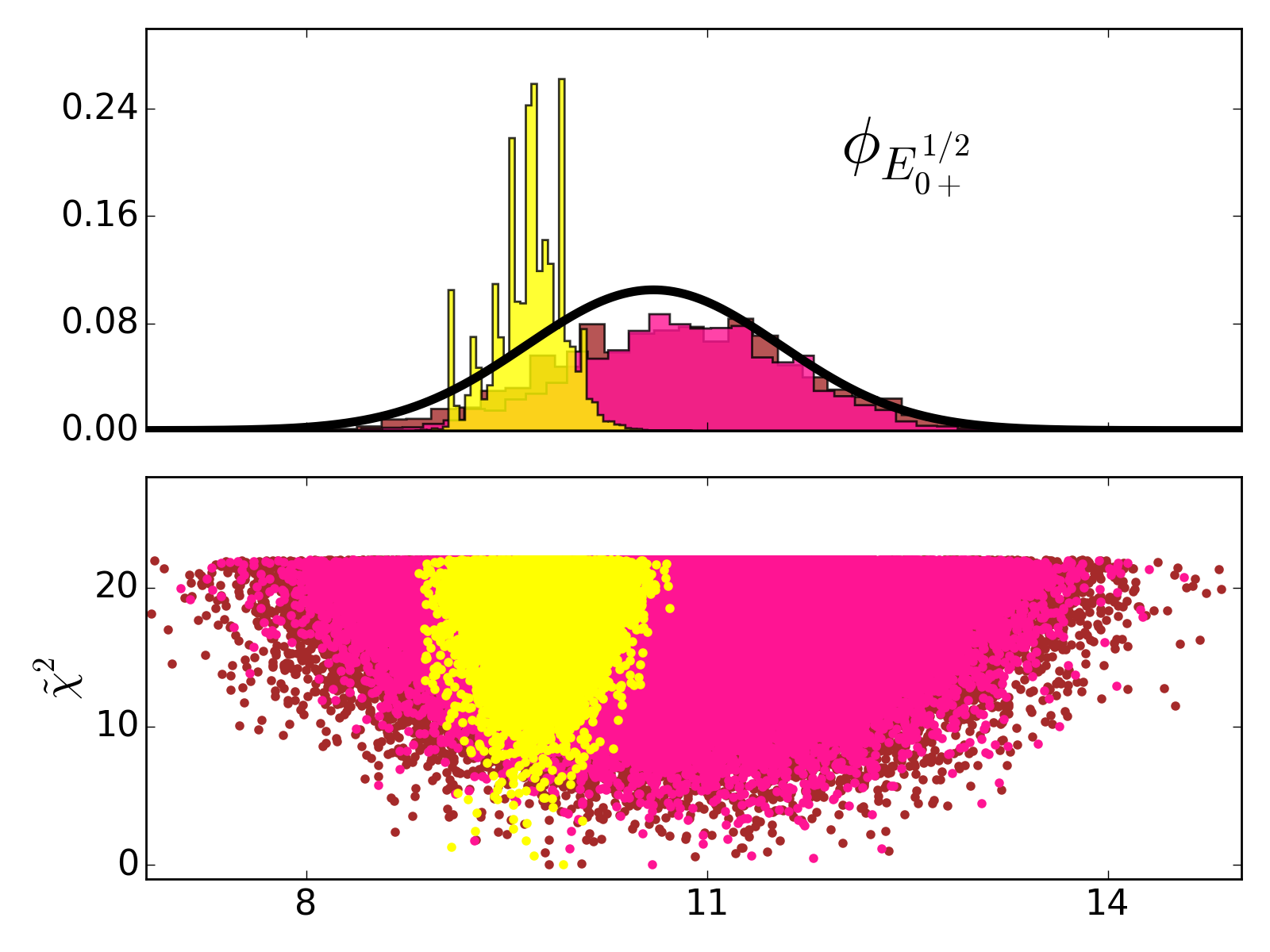}} \\[-1ex]
\caption{Top: Probability Distribution Functions normalized for the phase of $E_{0+}^{1/2}$  derived from set A (brown), 
set B (magenta) and set C (yellow). Phases are given in degrees. 
The black Normal is the distribution used to sample the MC space. 
Bottom: Scatter plot of $\tilde{\chi}^2$ VS $\phi_{E_{0+}^{1/2}}$ 
where  $\tilde{\chi}^2 = \chi^2-\chi^2_{min}$.} 
\label{fig:fw_paper_phase_and_chi2vsep0ph1}
\end{figure}

For the analyses with ``Varied'' phases known, Normal distributions  were utilized for the phase randomization. 
The analyses of pseudodata sets A and B yield phases nearly identical to the 
 Normal distributions used for the phase variation.  
This indicates that  data of such precision do not  exhibit sensitivity to the magnitude of the phase variation we imposed. 
 The derived phases from set C, the most precise analyzed  pseudodata set, emerge significantly narrower than the   Normal distributions 
 utilized to vary them. Fig. \ref{fig:fw_paper_phase_and_chi2vsep0ph1} shows 
  the PDF of the $E_{0+}^{1/2}$ phase derived from each pseudodata set. The phase PDFs derived from set A and B exactly match 
  the distribution used to vary the phase and which is marked by a black continuous curve. The  
   $E_{0+}^{1/2}$ phase derived  from set C emerges much narrower. 

   \afterpage{%
    \clearpage
    \thispagestyle{empty}
    \begin{landscape}
        \centering 
     
\resizebox{1.4\textwidth}{!}{\begin{tabular}{l|l|l|l|l||l|l|l||l|l|l}

    \hline
      \multicolumn{2}{c|}{Dataset} &
      \multicolumn{3}{c||}{Set A} &
      \multicolumn{3}{c||}{Set B} &
      \multicolumn{3}{c}{Set C} \\
      \hline
    Multipole     &   Generator    &              MD07             &              WI08             &            Varied             &             MD07              &             WI08              &            Varied             &             MD07              &             WI08             &             Varied            \\
    \hline
    \hline      
$E_{0+}^{1/2}$    &         6.582  &       6.87  $\pm$       0.21  &       6.88  $\pm$       0.21  &       6.88  $\pm$       0.21  &       6.46  $\pm$       0.12  &       6.48  $\pm$       0.11  &       6.48  $\pm$       0.12  &      6.624  $\pm$      0.019  &      6.638  $\pm$      0.019  &      6.616  $\pm$      0.024 \\
$E_{1+}^{1/2}$    &         1.263  &       1.29  $\pm$       0.08  &       1.28  $\pm$       0.08  &       1.28  $\pm$       0.08  &       1.24  $\pm$       0.04  &       1.23  $\pm$       0.04  &       1.22  $\pm$       0.04  &      1.262  $\pm$      0.007  &      1.253  $\pm$      0.007  &      1.263  $\pm$      0.007 \\
$E_{2+}^{1/2}$    &         0.361  &       0.30  $\pm$       0.05  &       0.30  $\pm$       0.05  &       0.30  $\pm$       0.05  &       0.42  $\pm$       0.03  &       0.41  $\pm$       0.03  &       0.41  $\pm$       0.03  &      0.355  $\pm$      0.004  &      0.349  $\pm$      0.004  &      0.355  $\pm$      0.004 \\
$E_{2-}^{1/2}$    &         2.119  &       2.05  $\pm$       0.15  &       2.03  $\pm$       0.15  &       2.03  $\pm$       0.15  &       2.20  $\pm$       0.08  &       2.18  $\pm$       0.08  &       2.17  $\pm$       0.08  &      2.108  $\pm$      0.013  &      2.093  $\pm$      0.012  &      2.109  $\pm$      0.015 \\
$M_{1+}^{1/2}$    &         1.804  &       1.43  $\pm$       0.37  &       1.43  $\pm$       0.36  &       1.45  $\pm$       0.37  &       1.77  $\pm$       0.20  &       1.83  $\pm$       0.19  &       1.74  $\pm$       0.24  &      1.829  $\pm$      0.032  &      1.896  $\pm$      0.031  &      2.043  $\pm$      0.071 \\
$M_{2+}^{1/2}$    &         0.260  &       0.28  $\pm$       0.11  &       0.29  $\pm$       0.12  &       0.29  $\pm$       0.12  &       0.30  $\pm$       0.06  &       0.30  $\pm$       0.06  &       0.30  $\pm$       0.06  &      0.254  $\pm$      0.010  &      0.254  $\pm$      0.009  &      0.256  $\pm$      0.011 \\
$M_{1-}^{1/2}$    &         1.575  &       1.89  $\pm$       0.24  &       1.90  $\pm$       0.24  &       1.92  $\pm$       0.24  &       1.54  $\pm$       0.13  &       1.55  $\pm$       0.12  &       1.53  $\pm$       0.13  &      1.575  $\pm$      0.021  &      1.575  $\pm$      0.020  &      1.571  $\pm$      0.023 \\
$M_{2-}^{1/2}$    &         0.560  &       0.35  $\pm$       0.10  &       0.34  $\pm$       0.10  &       0.35  $\pm$       0.11  &       0.59  $\pm$       0.05  &       0.59  $\pm$       0.05  &       0.58  $\pm$       0.05  &      0.554  $\pm$      0.008  &      0.551  $\pm$      0.008  &      0.557  $\pm$      0.009 \\
$E_{0+}^{3/2}$    &        12.624  &       11.9  $\pm$        0.5  &      12.0   $\pm$       0.5   &      12.0   $\pm$       0.5   &      12.84  $\pm$       0.28  &      12.94  $\pm$       0.27  &      12.90  $\pm$       0.28  &     12.584  $\pm$      0.047  &     12.694  $\pm$      0.044  &     12.689  $\pm$      0.060 \\
$E_{1+}^{3/2}$    &         0.697  &       0.71  $\pm$       0.15  &       0.72  $\pm$       0.15  &       0.70  $\pm$       0.17  &       0.64  $\pm$       0.08  &       0.65  $\pm$       0.08  &       0.65  $\pm$       0.09  &      0.689  $\pm$      0.013  &      0.703  $\pm$      0.014  &      0.663  $\pm$      0.021 \\
$E_{2+}^{3/2}$    &         0.549  &       0.67  $\pm$       0.11  &       0.67  $\pm$       0.11  &       0.67  $\pm$       0.11  &       0.47  $\pm$       0.05  &       0.47  $\pm$       0.05  &       0.46  $\pm$       0.05  &      0.544  $\pm$      0.009  &      0.540  $\pm$      0.008  &      0.542  $\pm$      0.009 \\
$E_{2-}^{3/2}$    &         4.665  &       4.71  $\pm$       0.32  &       4.73  $\pm$       0.32  &       4.73  $\pm$       0.32  &       4.76  $\pm$       0.18  &       4.76  $\pm$       0.18  &       4.75  $\pm$       0.18  &      4.637  $\pm$      0.031  &      4.649  $\pm$      0.030  &      4.640  $\pm$      0.034 \\
$M_{1+}^{3/2}$    &        36.162  &      36.33  $\pm$       0.14  &      36.30  $\pm$       0.15  &      36.26  $\pm$       0.16  &      36.05  $\pm$       0.08  &      36.02  $\pm$       0.08  &      36.05  $\pm$       0.09  &     36.160  $\pm$      0.021  &     36.129  $\pm$      0.021  &     36.095  $\pm$      0.021 \\
$M_{2+}^{3/2}$    &         0.212  &       0.15  $\pm$       0.30  &       0.10  $\pm$       0.30  &       0.15  $\pm$       0.27  &       0.27  $\pm$       0.14  &       0.26  $\pm$       0.13  &       0.24  $\pm$       0.14  &      0.231  $\pm$      0.025  &      0.208  $\pm$      0.023  &      0.230  $\pm$      0.031 \\
$M_{1-}^{3/2}$    &         6.576  &       7.22  $\pm$       0.46  &       7.23  $\pm$       0.44  &       7.29  $\pm$       0.46  &       6.91  $\pm$       0.23  &       6.96  $\pm$       0.23  &       6.93  $\pm$       0.23  &      6.566  $\pm$      0.037  &      6.604  $\pm$      0.037  &      6.579  $\pm$      0.040 \\
$M_{2-}^{3/2}$    &         0.539  &       0.94  $\pm$       0.19  &       0.92  $\pm$       0.19  &       0.92  $\pm$       0.20  &       0.45  $\pm$       0.10  &       0.45  $\pm$       0.10  &       0.45  $\pm$       0.09  &      0.546  $\pm$      0.016  &      0.542  $\pm$      0.015  &      0.550  $\pm$      0.017 \\
   \hline
    \end{tabular}}
      \captionof{table}{\label{tab:fw_paper_tab_moduli} Moduli of all extracted multipoles from single-energy analyses  
of pseudodata with relative, in comparison to that of existing data, uncertainty $2.0$ (set A), $1.0$ (set B) and $0.16$ (set C). Columns labeled 
``MD07'' refer to analyses with the multipole phases fixed to the MAID07 model values while columns ``WI08'' to analyses 
 with the multipole phases fixed to the WI08 values. Column ``Varied'' refers to analyses where  
 multipole phases were Gaussianly varied with $N[\mu, 2\sigma]$ 
where the mean value is from the $WI08$ solution and the uncertainty from the the $WI08$ single energy fit.
Multipoles given in $10^{-3}/m_{\pi^{+}}$ units.}
     \end{landscape}
    \clearpage
}


\section{Example: Application to the Bates and Mainz data at $Q^2=0.127$ $(GeV/c)^2$}
\label{sec:fw_bates}
The methodology of Sec. \ref{sec:fw_methodology}  was applied for a re-analysis of the $H(e,e',p)\pi^{0}$ 
Bates/Mainz measurements performed at $Q^2=0.127$ $GeV^2/c^2$ and $W=1232$ $MeV$. 
 The detailed description and analysis of this data can be found in ref. \cite{sparveris2005investigation}. 
 The data set consists of cross section results  for  $\sigma_{TT}$,  $\sigma_{LT}$, $\sigma_{0}$, $\sigma_{E2}$
 and the polarized beam cross section $\sigma_{LT'}$. The observables are defined as in ref. \cite{dreschsel1992threshold}. 
 
 As the data concern  $\gamma p \to p \pi^0$ measurements, 
 model input (MAID07) was used   to allow  the isospin separation of 
  multipoles  and only  few multipoles were derived.  
 The derived parameters are the $l=0$ charge multipole amplitudes (the $A_{p \pi^{0}}$ multipoles of Eq. \ref{eq:epjaiso}) 
 and the $l=1$ resonant multipole amplitudes with 
 isospin $I=3/2$. The $I=1/2$ multipoles were  fixed 
 to the MAID07 model values.  We performed two new analyses of the data: in the first, the multipole phases were fixed to the 
 $\pi N$ values \cite{gwuweb2}; in the  second the  $P_{33}$ phase was varied with Gaussian weight, 
 with mean value the $\pi N$ scattering 
 phase shift value and five times the experimental standard deviation ($\sigma$) of the scattering  phase shift. 
 The derived multipoles, which are listed in Table 
 \ref{tab:fw_paper_tab_ampli_bates}, were (statistically) identical in both cases; 
 the phase variation  did not induce any changes to the derived multipole amplitudes. Our results are in good agreement with 
 earlier analyses of the same data  \cite{sparveris2005investigation,stiliaris2007multipole}. The 
 Electric-to-Magnetic and Coulomb-to-Magnetic ratios, EMR and CMR respectively, are also given. These are defined as 
 $EMR= E_{1+}^{3/2} / M_{1+}^{3/2}$ and $CMR= E_{1+}^{3/2} / S_{1+}^{3/2}$, 
 where the Coulomb multipole 
 $S_{1+}^{3/2}$ is connected to the longitudinal multipole, the photon's momentum $q$ and 
 the photon's energy $\omega$ through the relation $S_{1+}^{3/2} = \frac{\vec{q}_{cm}}{\omega_{cm}} L_{1+}^{3/2}$. EMR and CMR 
 serve as the accepted gauge of the magnitude of the deformation of the proton \cite{papanicolas2007shapes}.

\begin{table} [ht]
\caption{Moduli of extracted amplitudes from the Bates/Mainz data. Results are with the phases fixed to $\pi N$ values \cite{gwuweb2} 
and with a $5\sigma$ Gaussian variation of the $P_{33}$ phase.}
\label{tab:fw_paper_tab_ampli_bates}
\resizebox{\columnwidth}{!}{%
\begin{tabular}{l|l|l||l}
    \hline
    Multipole   &   Fixed     &   Varied & Ref. \cite{sparveris2005investigation} \\  
    \hline
    \multicolumn{4}{l}{$I^{3/2}$ multipole amplitudes} \\
    \hline
    $M_{1+}$   & 40.0 $\pm$ 0.8 & 40.0  $\pm$ 0.8 &$41.4 \pm 0.3$\\
    $E_{1+}$   &  1.1 $\pm$ 0.3 &  1.1  $\pm$ 0.3 &$0.95 \pm 0.12$\\
    $L_{1+}$   & 1.08 $\pm$ 0.13 & 1.09 $\pm$ 0.13 &$1.26\pm0.08$\\
    \hline
    \multicolumn{4}{l}{Reaction channel multipole amplitudes} \\
    \hline
    $E_{0+}$   &  3.3 $\pm$ 1.0 & 3.2 $\pm$ 1.0 &2.9\\
    $L_{0+}$   &  1.5 $\pm$ 0.5 & 1.6 $\pm$ 0.5 &2.3\\
    \hline
    \multicolumn{4}{l}{Ratios} \\ \hline
    $EMR(\%)$   &    $-2.8 \pm 0.8$ & $-2.8 \pm 0.8$ & $-2.3 \pm 0.3$ \\
    $CMR(\%)$   &    $-5.5 \pm 0.7$ & $-5.5 \pm 0.7$ & $-6.1 \pm 0.2$ \\
   \hline 
    \end{tabular}
    }
\end{table}

\section{Summary and Conclusions}
\label{sec:fw_conclusions}
Using the AMIAS methodology we explored the possible influence of the use of the Fermi-Watson theorem 
in pion photoproduction analyses. The current practice of using fixed (with no uncertainty) values was examined 
and compared to analyses where the $\pi  N$ phase values and their uncertainty were used as prior knowledge. 
The AMIAS was used for the first time to allow   prior knowledge  to be incorporated into experimental 
analyses.
Sets of pseudodata  of increasing statistical precision were analyzed and their multipole content was derived. 
In the case of  pseudodata of comparable statistical precision to the most recent pion photoproduction data 
the derived multipoles emerged nearly identical in mean value and uncertainty.  
The experimental phase uncertainty induced significant changes in the  derived multipole amplitude PDFs when the analyzed pseudodata 
 were created with precision six times  the statistical precision of current experimental data. 
 
 The same methodology was applied to the 
 $H(e,e',p)\pi^{0}$  Bates/Mainz data measured at  $Q^2=0.127$ $GeV^2/c^2$ and $W=1232$ $MeV$ where 
 even a $5\sigma$ phase variation of the experimentally derived phase values 
 did not induce any changes to the  derived multipoles. We conclude that 
 for the  current precision of pion photo-and electroproduction data the $\pi N$ phases taken from pion-nucleon 
 scattering as perfectly known 
 is justified. However, for the new generation of data aspiring to distinguish among different models of nucleon structure, 
 the type of analysis presented here where the experimentally derived phases are allowed to vary will need to be implemented. 
 
\section*{Acknowledgments}
\label{sec:acknowledgements}
This  work,  part  of  L.  Markou  Doctoral  Dissertation,  was
supported by the Graduate School of The Cyprus Institute.

\bibliographystyle{unsrt}
\bibliography{bibliography}
\end{document}